\documentclass[aps,prl,twocolumn,floatfix,superscriptaddress,amsmath,amssymb]{revtex4-1}

\usepackage{graphicx}
\usepackage{bm}
\usepackage{hyperref}
\usepackage[mathlines]{lineno}
\usepackage{braket}
\usepackage{bbm}
\usepackage{verbatim}
\usepackage{ulem}
\usepackage{standalone}
\usepackage{comment}
\usepackage[height=8.85in,width=6.45in]{geometry}
\usepackage{slashed}
\usepackage{braket}
\usepackage{tikz}
\usepackage{tikz-cd}
\usepackage{times}
\usepackage{courier}
\usepackage{mdframed}
\usepackage{datetime}
\usepackage{dsfont}

\usepackage{cancel}
\usepackage{caption}
\usepackage{subcaption}
\usepackage[compat=1.1.0]{tikz-feynman}
\usepackage{pifont}
\usepackage{array} 
\usepackage{mathtools}

\newcommand{\parR}[1]{\textbf{\textit{#1}}---}

\newcommand{\wid}[1]{\widehat{#1}}
\newcommand{\wZ}{\widehat{\mathbb{Z}}}

\renewcommand\[{\begin{equation}}
\renewcommand\]{\end{equation}}

\def\ie{\begin{equation}\begin{aligned}}
\def\fe{\end{aligned}\end{equation}}

\begin{document}

\title{Understanding deconfined quantum critical points from crystalline categorical Landau paradigm}	

\author{Hiromi Ebisu}
\email[Electronic address:$~~$]{hiromi.ebisu@riken.jp}
\affiliation{Interdisciplinary Theoretical and Mathematical Sciences Program (iTHEMS) RIKEN, Wako
351-0198, Japan}

\author{Bo Han}
\email[Electronic address:$~~$]{bhan2@thp.uni-koeln.de}
\affiliation{Institute for Theoretical Physics, University of Cologne, Zülpicher Str. 77, 50937 Cologne, Germany}

\author{Weiguang Cao}
\email[Electronic address:$~~$]{njua@outlook.com}
\affiliation{Center for Quantum Mathematics at IMADA, Southern Denmark University, Campusvej 55, 5230 Odense, Denmark}
\affiliation{Department of Physics, Hong Kong University of Science and Technology, Clear Water Bay, Hong Kong, China}
\affiliation{Center for Theoretical Condensed Matter Physics, Hong Kong University of Science and Technology, Clear Water Bay, Hong Kong, China}

\date{\today}



\begin{abstract}
Deconfined quantum critical points (DQCPs) involving lattice symmetries evade the conventional Landau paradigm because the competing orders break incompatible internal and crystalline symmetries. We show that a class of DQCPs can nevertheless be understood as Landau-type transitions after gauging anomalous onsite symmetries. For spin chains with Lieb–Schultz–Mattis (LSM) anomalies, gauging produces a noninvertible lattice translation whose fusion closes only up to ordinary translations, giving rise to a crystalline categorical symmetry. In the gauged description, the original DQCP becomes a transition between different symmetry-breaking patterns of this categorical symmetry. We demonstrate this mechanism in microscopic lattice models -- the magnetic–valence-bond-solid (VBS) DQCP realizes a $\mathrm{Rep}$($D_8$)-type crystalline categorical Landau transition, whereas a y-antiferromagnetic–VBS DQCP realizes a $\mathrm{Rep}$($H_8$)-type one. Although $\mathrm{Rep}$($D_8$) and $\mathrm{Rep}$($H_8$) share the same fusion rules, they have inequivalent $F$-symbols and therefore define distinct categorical descriptions. Our results show that the universal categorical structure underlying these DQCPs is encoded in the full fusion category, rather than in the fusion ring alone.
\end{abstract}
\maketitle

\parR{Introduction.} 
Deconfined quantum critical points (DQCPs) provide a paradigmatic example of quantum phase transitions beyond the conventional Landau paradigm, describing continuous transitions between phases that break incompatible symmetries
~\cite{senthil2004deconfined,senthil2004quantum,senthil2024deconfined}. 
The original motivation arose from the N\'eel–valence-bond-solid (VBS) transition in quantum magnets, where antiferromagnetic and VBS orders break unrelated spin-rotation and lattice 
symmetries. It was later understood that such transitions are closely related to Lieb–Schultz–Mattis (LSM) anomalies intertwining internal and crystalline symmetries~\cite{Lieb1961,metlitski2018intrinsic}.
Recently, simpler realizations of DQCPs have been identified in~$(1+1)$d spin systems, including continuous transitions between ferromagnetic (FM) and VBS phases~\cite{jiang2019ising,xi2022dynamical}. These models provide a useful microscopic arena for understanding the structure of DQCP and its relation to anomalies and dualities.
\par

On a different front, recent developments have revealed that gauging internal symmetries in LSM-anomalous lattice systems can generate generalized crystalline symmetries, including spatially modulated symmetries~\cite{,Pace:2025hpb,10.21468/SciPostPhys.20.4.117}
 and non-invertible lattice translations~\cite{Seifnashri:2023dpa,oishi2026non}. Unlike constructions where an internal non-invertible symmetry is imposed microscopically~\cite{PhysRevLett.98.160409,Cao:2024qjj,Cao:2025qnc,Cao:2025qhg,Hsin:2024aqb,Hsin:2025ria}, here the emergent symmetries are \textit{intrinsically crystalline}: their fusion closes only up to ordinary lattice translations~\cite{Seiberg:2023cdc,Cao:2023doz,Seiberg:2024gek,Gorantla:2024ocs}. This suggests that LSM anomalies may naturally provide a microscopic origin for categorical symmetries in lattice systems.

\begin{figure}[t]
    \centering
\includegraphics[scale=0.65]{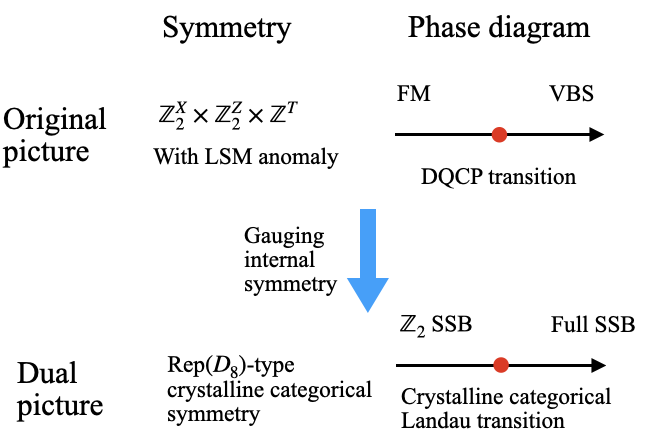}
    \caption{After gauging the internal symmetry, the DQCP with LSM anomalous symmetry is mapped to a crystalline categorical Landau transition.
    }
    \label{fig:landau}
\end{figure}

%
Motivated by this perspective, we investigate the dual descriptions of DQCPs obtained by gauging internal symmetries in concrete LSM-anomalous lattice systems.~(See, for instance,~\cite{PhysRevLett.130.026801,su2023boundary,PhysRevB.109.245108} for recent discussion of interpretation of the DQCP by dual description in a different setting.). 
We show that the $(1+1)$d FM–VBS DQCP can be mapped to a crystalline categorical Landau transition governed by a non-invertible lattice translation symmetry with an emergent Rep($D_8$)-type structure in the infrared (IR), as shown in Fig.~\ref{fig:landau}. In this dual description, the original DQCP is reinterpreted as 
a transition point between complete and partial symmetry-breaking phases of the crystalline categorical symmetry. 

We next show that the DQCP between the y-antiferromagnetic (yAFM) and VBS phases admits a distinct categorical Landau description, associated with a crystalline categorical symmetry of Rep($H_8$) type. Although Rep($D_8$) and Rep($H_8$) have the same fusion rules, they possess inequivalent $F$-symbols, and hence define distinct fusion categories. This distinction demonstrates that the categorical Landau descriptions of DQCPs are sensitive to the full fusion-category data, not merely the fusion ring. While related categorical perspectives on DQCPs and non-invertible symmetry breaking were recently discussed in Ref.~\cite{Chen:2025ivo}, our work realizes these structures in concrete microscopic models with LSM constraints via gauging anomalous symmetries.




\parR{Crystalline categorical Landau transition from DQCP in $(1+1)$d} 
Let us outline our general formalism 
to understand the DQCP via categorical Landau transition in a $(1+1)d$
model. (Similar ideas relating gauging and symmetry-breaking transitions in different models have appeared in e.g.,~\cite{PhysRevB.108.075105,10.21468/SciPostPhys.20.5.134,tmvy-vsqd}.) 
We start from the Hamiltonian on an Ising chain
\begin{align}
    \begin{split}
      H=&-J_x\sum_{j}\sigma^x_{j}\sigma^x_{j+1}-J_z\sum_{j}\sigma^z_{j}\sigma^z_{j+1}\\
      &+h_x\sum_{j}\sigma^x_{j}\sigma^x_{j+2}+h_z\sum_{j}\sigma^z_{j}\sigma^z_{j+2},  
    \end{split}
\end{align}
which has a robust second order phase transition from the zFM phase to the VBS phase~\cite{jiang2019ising}. 
Throughout, we assume the length of the chain is even.
The whole transition has the internal $\mathbb Z_2^z\times \mathbb Z_2^x$, generated by
\begin{equation}
    U_z=\prod_{j}\sigma^z_{j},\quad U_x=\prod_{j}\sigma^x_{j}
\end{equation}
together with an LSM anomaly with one site lattice translation symmetry $T$. The FM phase spontaneously breaks the $\mathbb Z_2^x$ internal symmetry, while the VBS phase spontaneously breaks the 
translational symmetry. 

Because of the LSM anomaly, after gauging the whole internal symmetry $\mathbb Z_2^x\times \mathbb Z_2^z$, we will get a non-invertible lattice translation symmetry besides the dual internal $\mathbb Z_2\times \mathbb Z_2$ symmetry~\cite{Seifnashri:2023dpa,oishi2026non}. This can be shown by sequentially gauging the internal symmetries. To this end, we can first gauge the $\mathbb Z_2^z$ symmetry, which is equivalent to a Hadamard gate plus a Kramers-Wannier (KW) transformation
\begin{equation*}
    \sigma^z_{j}\to \sigma^z_{j}\sigma^z_{j+1},\quad \sigma^x_{j}\sigma^x_{j+1}\to \sigma^x_{j+1}. 
\end{equation*}
After gauging, the $\mathbb Z_2^x$ symmetry becomes a dipole symmetry generated by $U_d=\prod_{j}(\sigma^x_j)^j$. 
This mirrors the fact that when we gauge subgroup of the LSM system, a modulated symmetry emerges~\cite{Pace:2025hpb,10.21468/SciPostPhys.20.4.117}. To proceed, 
we gauge the remaining dipole symmetry, with the following spatially modulated Gauss law 
\begin{equation}\label{gauss}
    G_j=\tau^z_{j-\frac{1}{2}}(\sigma^x_j)^j\tau^z_{j+\frac{1}{2}}.
\end{equation}
Now the one site lattice translation is not gauge invariant as $ TG_{j}\neq G_{j+1}T$~\footnote{However, the square of lattice translation is gauge invariant $ T^2G_{j}= G_{j+1}T^2$ and is a valid symmetry in the dual models.}. To remedy this issue, 
we define the Kennedy-Tasaki (KT) transformation~\cite{kennedy1992hidden,Li:2023ani} with both $\sigma$ and $\tau$ variables as
\begin{align}
    \begin{split}
      \mathcal{D}_{\text{KT}}:\quad &\tau^z_{j-\frac{1}{2}}\sigma^x_j\tau^z_{j+\frac{1}{2}}\leftrightarrow \tau^z_{j-\frac{1}{2}}\tau^z_{j+\frac{1}{2}},  \\
      &\sigma^z_j\tau^x_{j+\frac{1}{2}}\sigma^z_{j+1}\leftrightarrow \sigma^z_j\sigma^z_{j+1}.
    \end{split}
\end{align}
The combination of KT and lattice translation $\mathcal{T}=\mathcal{D}_{\text{KT}}T$ is therefore gauge invariant
\begin{equation*}
    \mathcal{T}G_{j}= G_{j+1}\mathcal{T},
\end{equation*}
but non-invertible
\begin{equation*}
    \mathcal{T}\times \mathcal{T}=\left(1+\prod_{j}\sigma^x_j\right)\left(1+\prod_{j}\tau^x_j\right)T^2,
\end{equation*}
which comes from the fusion rule of the KT transformation. Because the fusion of non-invertible translation $\mathcal{T}$ includes ordinary lattice translation, this dual symmetry does not form a standard fusion category~\cite{oishi2026non}. In next section, we show that this non-invertible lattice translation $\mathcal{T}$ has a Rep($D_{8}$)-type structure in the low energy effective theory.

Because discrete gauging will keep the phase diagram structure, the original DQCP can be mapped and analyzed in the dual picture. After gauging, the dual FM phase spontaneously breaks a single dual $\mathbb Z_2$ internal symmetry, while the dual VBS phase breaks the full crystalline categorical symmetry.  Therefore, the whole DQCP will be mapped to a \textit{Landau-type transition}, i.e. a transition from a larger symmetry breaking to a smaller symmetry breaking, with necessary inclusion of a non-invertible lattice translation symmetry. 

\parR{Concrete lattice demonstration of crystalline categorical Landau transition}
We have sketched the symmetry and gauging procedure of the standard FM-VBS DQCP. To identify the symmetry breaking patterns after gauging, it is sufficient to focus on a representative transition of two models with exact ground states
\begin{equation}\label{eq:dqcp}
    H=(1-\lambda)H_{\text{FM}}+\lambda H_{\text{VBS}},
\end{equation}
where
\begin{equation}
    H_{\text{FM}}=-\sum_{j}\sigma^z_{j}\sigma^z_{j+1},\label{9}
\end{equation}
realizes the FM phase with ground states spanned by $\ket{00\cdots 0},\ket{11\cdots 1}$ that break $\mathbb{Z}_2^x$ symmetry, and 
\begin{align}
    \begin{split}
       H_{\text{VBS}}=&h_x'\sum_{j}(I-\sigma^x_{j}\sigma^x_{j+1})(I-\sigma^x_{j+1}\sigma^x_{j+2})\\
       &+h_z'\sum_{j}(I-\sigma^z_{j}\sigma^z_{j+1})(I-\sigma^z_{j+1}\sigma^z_{j+2}), 
    \end{split}\label{10}
\end{align}
leads to the VBS phase with ground states spanned by
\begin{eqnarray}
  \ket{GS}_1=  \bigotimes_{k}(\ket{00}_{2k-1,2k}+\ket{11}_{2k-1,2k}),\nonumber\\\ket
{GS}_2=\bigotimes_{k}(\ket{00}_{2k,2k+1}+\ket{11}_{2k,2k+1}),\nonumber
\end{eqnarray}
which preserve the internal symmetry but spontaneously breaks the lattice translation symmetry. 
Preliminary numerical results~\cite{jiang2019ising,PhysRevB.99.165143} suggest that a DQCP occurs by tuning the  parameter $\lambda$. 

\paragraph{(i)~FM}
We turn to dual description of these two models~\eqref{9}\eqref{10}. 
We first consider the FM phase. After gauging $U_z$, we have
\begin{eqnarray}
    H_{\text{FM}}^\prime= -\sum_{j=1}^{L}\sigma^z_{j-1}\sigma^z_{j+1}\label{fm2}.
\end{eqnarray}
Then, we gauge the residue dipole symmetry with a gauged Hamiltonian~\eqref{fm2} as
\begin{eqnarray}
    H^{\prime\prime}_{\text{FM}}=-\sum_{j}\sigma^z_{j-1}(\tau^x_{j-\frac{1}{2}})^{j-1}(\tau^x_{j+\frac{1}{2}})^{j-1}\sigma^z_{j+1}\label{fm3}
\end{eqnarray}
so that it commutes with the spatially modulated Gauss law~\eqref{gauss}. To proceed, we apply the following unitary transformation on the Hamiltonian
\begin{equation}\label{eq:unitary}
    \mathcal{U}_{\text{CZ}}=\prod_{j}\text{CZ}_{j-\frac{1}{2},j}\text{CZ}_{j,j+\frac{1}{2}}.
\end{equation}
It is product of  controlled Z gates on adjacent sites and links.
The Hamiltonian becomes
\begin{eqnarray}
   {H}_{\text{FM}}^{\prime\prime\prime}=-\sum_{k=1}\left(\sigma^z_{2k-2}\sigma^z_{2k}+\tau^x_{2k-\frac{1}{2}}\tau^x_{2k+\frac{1}{2}}\right),\label{fm4}
\end{eqnarray}
whereas the Gauss law~\eqref{gauss} is simplified to
\begin{eqnarray}\label{eq:simplegauss}
  \sigma^x_{2k-1}=1,\quad \tau^z_{2k-\frac{1}{2}}\tau^z_{2k+\frac{1}{2}}=1.\label{gauss2}
\end{eqnarray}
The first condition in~\eqref{gauss2} decouples the odd-site $\sigma$ spins and we define $\widehat{\sigma}^{z(x)}_{k}=\sigma^{z(x)}_{2k}$. The second condition in~\eqref{gauss2} indicates that we can group two adjacent $\tau$ spins, e.g. $\tau_{2k-\frac{1}{2}}$ and $\tau_{2k+\frac{1}{2}}$, into one unit cell, and effectively two spins in the same unit cell have to be aligned in the same direction, allowing us to 
introduce effective Pauli operators $\widehat{\tau}^z_{k},\widehat{\tau}^x_{k}$ acting on the $k$-th unit cell.
\par
In this effective description, the dual internal symmetry becomes $\widehat{\mathbb Z}_2^{\sigma}\times \widehat{\mathbb Z}_2^{\tau}$, 
generated by
\begin{equation}\label{eq:effectivesym}
    \widehat{U}_{\sigma}=\prod_{k}\widehat{\sigma}^x_{k},\quad  \widehat{U}_{\tau}=\prod_{k}\widehat{\tau}^x_{k}.
\end{equation}
The non-invertible lattice translation now transforms the operators as
\begin{align*}
    \begin{split}
          \widehat{\mathcal{T}}:\quad &\widehat{\sigma}^x_{k}\to \widehat{\tau}^z_{k}\widehat{\tau}^z_{k+1},\quad  \widehat{\tau}^z_{k}\widehat{\tau}^z_{k+1}\to \widehat{\sigma}^x_{k+1},\\
          &\widehat{\tau}^x_{k}\to \widehat{\sigma}^z_{k}\widehat{\sigma}^z_{k+1},\quad  \widehat{\sigma}^z_{k}\widehat{\sigma}^z_{k+1}\to \widehat{\tau}^x_{k+1}.
    \end{split}
\end{align*}
implying that the corresponding operator is $\widehat{\mathcal{T}}=S\widehat{D}_\sigma \widehat{D}_{\tau}$, 
where $S$ denotes swap operator that exchanges $\widehat{\sigma}_k$ and $\widehat{\tau}_k$ and $\widehat{D}_{\sigma(\tau)}$ represents KW duality operator for $\widehat{\sigma}(\widehat{\tau})$ variables.  
The symmetry structure of the non-invertible translation along with the internal symmetry $\widehat{\mathbb Z}_2^{\sigma}\times \widehat{\mathbb Z}_2^{\tau}$, form a Rep($D_8$)-type crystalline categorical symmetry.

The fusion rule of $\widehat{\mathcal{T}}$ follows from its expression
\begin{equation}
    \widehat{\mathcal{T}}\times \widehat{\mathcal{T}}=(1+\widehat{U}_{\sigma})(1+\widehat{U}_{\tau})\widehat{T}_x^2,\label{fusionT}
\end{equation}
where $\widehat{T}_x$ denotes effective lattice translation that moves the $\widehat{\sigma},\widehat{\tau}$ spins by one unit cell.

\begin{figure}[t]
    \centering
\includegraphics[scale=0.35]{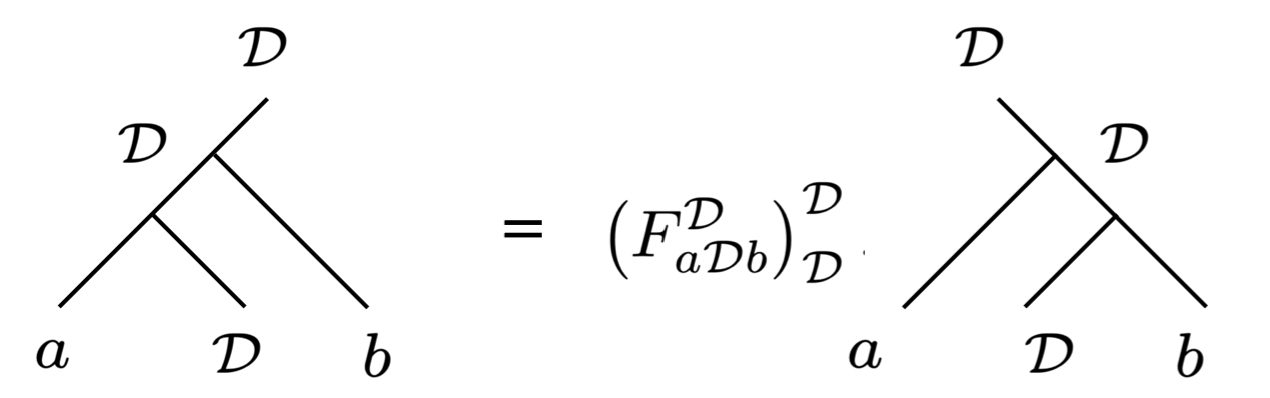}
    \caption{This $F$-symbol measures the relative sign between two fusion orders:\((a\otimes \mathcal D)\otimes b\) vs \(
a\otimes(\mathcal D\otimes b).
\)
    }
    \label{fig:fsymbol}
\end{figure} 
However, fusion rule alone does not fix the categorical symmetry. For example, Rep($D_8$) and Rep($H_8$) fusion categories share the same fusion rule, yet are distinct from each other in bicharacter~\cite{Bhardwaj:2017xup}~\footnote{Actually there is another data called Frobenius-Schur (FS) indicator, and there are in total four different TY($\mathbb Z_2\times \mathbb Z_2$) fusion categories. See~[33] for a detailed discussion. Here we only focus on Rep($D_8$) and Rep($H_8$).}, 
which is off-diagonal for \(\mathrm{Rep}(D_8)\) and diagonal for \(\mathrm{Rep}(H_8)\). This data is also reflected in $F$-symbols, for example,
\begin{equation}
    \left(F_{a\mathcal D b}^{\mathcal D}\right)^{\mathcal D}_{\mathcal D},\label{Fsym}
\end{equation}
which measures the relative sign between two fusion orders shown in Fig.~\ref{fig:fsymbol}. Here, \(a,b\in \mathbb Z_2\times \mathbb Z_2\)  and \(\mathcal D\) is the non-invertible operator, which is given by 
$S\widehat{D}_\sigma \widehat{D}_{\tau}$ in the present case. For crystalline categorical symmetries, we evaluate the $F$-symbol~\eqref{Fsym} by considering fusion of topological symmetry defects~\cite{PhysRevB.104.115156,Seifnashri:2023dpa} to distinguish Rep($D_8$)- and Rep($H_8$)-type. We leave the calculation details in the Appendix. 

\par
After identifying the dual categorical symmetry, we study 
the dual description of the FM Hamiltonian~\eqref{fm4}
\begin{eqnarray}
    \widehat{H}_{\text{FM}}=-\sum_{k}\left(\widehat{\sigma}^z_{k-1}\widehat{\sigma}^z_{k}+\widehat{\tau}^x_{k}\right),\label{fm5}
\end{eqnarray}
which has two ground states
\begin{align}
    \begin{split}
       &\ket{GS}_1=\ket{FM+}_\sigma\otimes\ket{PM}_\tau,\\
       &\ket{GS}_2=\ket{FM-}_\sigma\otimes\ket{PM}_\tau.
    \end{split}
\end{align}
Here, $\ket{FM\pm}_\sigma=\ket{00\cdots 0}_{\sigma},\ket{11\cdots 1}_{\sigma}$ are ferromagnet states for $\widehat{\sigma}$ spin and $\ket{PM}_\tau=\ket{++\cdots +}_{\tau}$ is the paramagnet state for $\widehat{\tau}$ spin. Note that the non-invertible translation symmetry remains unbroken in the dual FM phase because it has a non-vanishing expectation value in every ground state
\begin{equation*}
\langle GS_i| \widehat{\mathcal{T}}|GS_i\rangle \neq 0,
\qquad i=1,2,
\end{equation*}
derived from its action on the ground states
\begin{align*}
  \widehat{\mathcal{T}} |GS\rangle_1 &= |GS\rangle_1+|GS\rangle_2, \\
  \widehat{\mathcal{T}} |GS\rangle_2 &= |GS\rangle_1+|GS\rangle_2.
\end{align*}
In contrast, the invertible symmetry $\wZ_2^\sigma$ is spontaneously broken, and the dual FM phase realizes a partial symmetry breaking of the full Rep$(D_8)$-type crystalline symmetry. 
\paragraph{(ii)~VBS}
Next, we investigate the dual description of the VBS phase. After gauging $\mathbb Z_2^z$, the Hamiltonian becomes
\begin{align}
    \begin{split}
       H_{\text{VBS}}^\prime=&h^\prime_x\sum_J(I-\sigma^x_{j})(I-\sigma^x_{j+1})\\
    &+h^\prime_z\sum_j(I-\sigma^z_{j-1}\sigma^z_{j+1})(I-\sigma^z_{j}\sigma^z_{j+2}).\label{VBS2} 
    \end{split}
\end{align}
We further gauge the dipole symmetry by considering the minimal coupled Hamiltonian
\begin{align}
    \begin{split}\label{vbs3}
        H_{\text{VBS}}^{\prime\prime}=&h^\prime_x\sum_j(I-\sigma^x_{j})(I-\sigma^x_{j+1})\\
        &+h^\prime_z\sum_j\left[I-\sigma^z_{j-1}(\tau^x_{j-1/2})^{j-1}(\tau^x_{j+1/2})^{j-1}\sigma^z_{j+1}\right]\\
        &\times\left[I-\sigma^z_{j}(\tau^x_{j+1/2})^{j}(\tau^x_{j+3/2})^{j}\sigma^z_{j+2}\right],
    \end{split}
\end{align}
together with the Gauss law~\eqref{gauss}. 
Similar to the FM case, after applying the unitary~\eqref{eq:unitary}, imposing the simplified Gauss law~\eqref{eq:simplegauss} and using the effective description, we have
\begin{align}
    \begin{split}
        \widehat{H}_{\text{VBS}}=
        & h^\prime_x\sum_{k}[\left(I-\widehat{\sigma}^x_{k}\right)\left(I-\widehat{\tau}^z_{k}\widehat{\tau}^z_{k+1}\right)\\
        &+\left(I-\widehat{\tau}^z_{k}\widehat{\tau}^z_{k+1}\right)\left(I-\widehat{\sigma}^x_{k+1}\right)]\\
        &+h^\prime_z\sum_{k}[\left(I-\widehat{\sigma}^z_{k-1}\widehat{\sigma}^z_{k}\right)\left(I-\widehat{\tau}^x_{k}\right)\\
        &+\left(I-\widehat{\tau}^x_{k}\right)\left(I-\widehat{\sigma}^z_{k}\widehat{\sigma}^z_{k+1}\right)].\label{vbs5}
    \end{split}
\end{align}
which has five ground states
\begin{eqnarray}\label{eq:vbsgs}
    \ket{GS~\pm,\pm}&=&\ket{FM\pm}_\sigma\otimes\ket{FM\pm}_\tau,\nonumber\\
    \ket{GS}_0&=&\ket{PM}_\sigma\otimes \ket{PM}_\tau.\label{27}
\end{eqnarray}
The action of non-invertible translation on the ground states is given by
\begin{align}
    \begin{split}
      \widehat{\mathcal{T}}\ket{GS~\pm,\pm}=&\ket{GS}_0,\\
      \widehat{\mathcal{T}}\ket{GS}_0=&\ket{GS~+,+}+\ket{GS~+,-}\\
    &+\ket{GS~-,+}+\ket{GS~-,-},
    \end{split}
\end{align}
which shows that the non-invertible translation does not preserve any individual ground state. Together with the spontaneous breaking of the dual internal symmetries, this implies that the dual VBS phase realizes complete spontaneous breaking of the Rep($D_8$)-type crystalline symmetry.

In summary, we map the original DQCP~\eqref{eq:dqcp} into a dual Landau type transition
\begin{equation}\label{eq:crystallandau}
    \widehat{H}=(1-\lambda)\widehat{H}_{\text{FM}}+\lambda \widehat{H}_{\text{VBS}}
\end{equation}
that connects a complete Rep$(D_8)$-type crystalline symmetry breaking phase and a partial $Z_2$ symmetry breaking phase, at the cost of extenbding ordinary symmetries by crystalline categorical symmetries.

\begin{figure}[t]
    \centering
\includegraphics[scale=0.35]{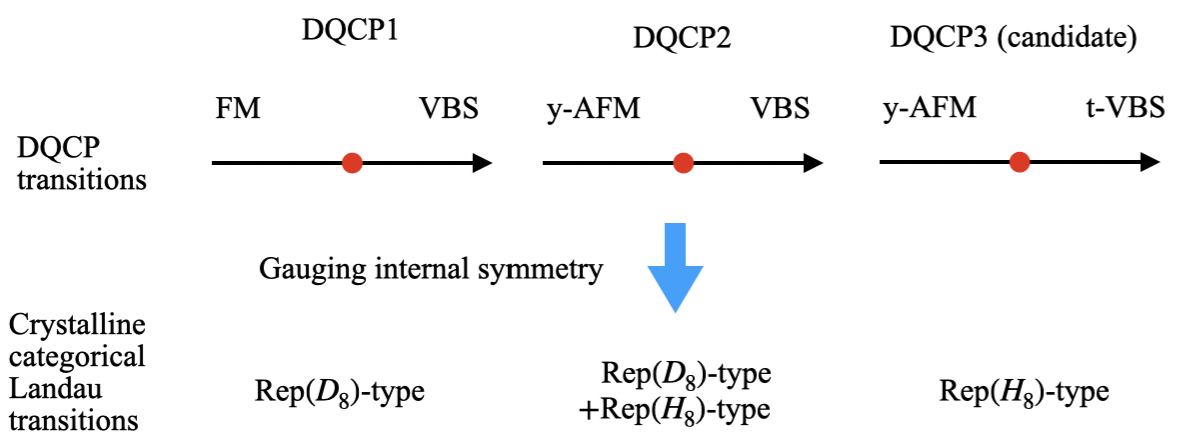}
    \caption{DQCPs with different crystalline categorical Landau origins. In particular, DQCP1 and DQCP2 are captured by Eq.~(\ref{eq:crystallandau}) and (\ref{eq:crystallandau2}), respectively, while DQCP3 is  briefly discussed in the Outlook.
    }
    \label{fig:DQCPs}
\end{figure} 
\parR{Rep($H_8$)-type crystalline categorical Landau transition}
The previous example demonstrates that gauging an LSM-anomalous system with ordinary lattice translation symmetry generates a Rep($D_8$)-type non-invertible translation.  
A natural question is whether other crystalline realizations of LSM anomalies can generate distinct categorical symmetries and therefore lead to different categorical Landau descriptions.

As another concrete example of the above question, we consider an LSM-anomalous system with a twisted lattice translation symmetry, in which the lattice translation acts nontrivially on the internal symmetry. Unlike the ordinary translation considered above, gauging such a system generates a non-invertible twisted translation whose categorical structure is governed by Rep($H_8$) rather than Rep($D_8$).
To realize this scenario concretely, we study a yAFM–VBS transition protected by a Hadamard-twisted translation symmetry. While the same transition can also be described using the ordinary lattice translation and therefore admits a Rep($D_8$)-type crystalline Landau description, we can in principle add deformation that only keeps the twisted-translation such that it will lead to a genuinely Rep($H_8$)-type crystalline categorical symmetry.

We start from an exotic LSM anomaly~\cite{Furukawa:2025flp} between internal
\(\mathbb Z_2^{YZ}\times \mathbb Z_2^{ZY}\) symmetry generated by
\begin{equation}
    U_{YZ}=\prod_k \sigma^y_{2k}\sigma^z_{2k+1},
    \qquad
    U_{ZY}=\prod_k \sigma^z_{2k}\sigma^y_{2k+1},\label{two}
\end{equation}
and the one-site translation symmetry \(T\) that exchanges the two internal symmetry
\begin{equation}
    T U_{YZ} T^{-1}=U_{ZY}.
\end{equation}
The LSM anomaly in the microscopic symmetry $ \left(\mathbb Z_2^{YZ}\times \mathbb Z_2^{ZY}\right)\rtimes \mathbb Z^T$ is obvious as we rewrite the internal symmetry operators as
\begin{equation}
    U_{YZ}=\prod_j a_j,
    \qquad
    U_{ZY}=\prod_j b_j,
\end{equation}
with the local representatives
\begin{equation}
    a_j=
    \begin{cases}
        \sigma^y_j, & j \ \text{even},\\
        \sigma^z_j, & j \ \text{odd},
    \end{cases}
    \qquad
    b_j=
    \begin{cases}
        \sigma^z_j, & j \ \text{even},\\
        \sigma^y_j, & j \ \text{odd}.
    \end{cases}
\end{equation}
This is because $a_j b_j=-b_j a_j~\forall j$, 
and each microscopic site carries a projective representation of
\(\mathbb Z_2^{YZ}\times \mathbb Z_2^{ZY}\).

If lattice translation $T$ flows in the IR to an internal order-two symmetry, the effective internal symmetry generated by \(U_{YZ}\), \(U_{ZY}\), and the infrared remnant of translation has the structure of \(D_8\), with an anomaly. The dual non-invertible symmetry is then expected to be of \(\mathrm{Rep}(H_8)\) type rather than \(\mathrm{Rep}(D_8)\) type~\cite{Diatlyk:2023fwf}. The distinction will be diagnosed by the lattice \(F\)-symbol.


For later convenience, we pass to a unitarily equivalent basis. First exchange \(\sigma^x\) and \(\sigma^y\) on every site and then exchange \(\sigma^x\) and \(\sigma^z\) only on odd sites. Since this is only a local unitary change of basis, the LSM anomaly is unchanged. In the new basis, the symmetry becomes
\begin{equation}\label{eq:secondsym}
    \left(\mathbb Z_2^x\times \mathbb Z_2^z\right)\rtimes \mathbb Z^{T'},
\end{equation}
generated by 
\begin{equation}
    U_x=\prod_j \sigma^x_j,
    \qquad
    U_z=\prod_j \sigma^z_j,\label{twoG}
\end{equation}
and the Hadamard-twisted translation (twisted translation for short)
\begin{equation}
    T'=T \prod_j H_j ,\label{twistedT}
\end{equation}
where \(H_j\) is the Hadamard gate on site \(j\).

To parallel the Rep($D_8$) construction above, we now use the second representation~\eqref{eq:secondsym} with twisted translation  and construct the corresponding symmetric lattice models.
In particular, we focus on two gapped phases, yAFM and the VBS phases. The former is 
described by 
\begin{eqnarray}
    H_{\text{yAFM}}=\sum_j \sigma^y_j\sigma^y_{j+1},\label{afm00}
\end{eqnarray}
whereas the latter is by the same Hamiltonian as~\eqref{10}.
The ground states of the yAFM phase spontaneously breaks diagonal subgroup of $\mathbb{Z}_2^x\times\mathbb{Z}_2^z$, 
whereas the VBS phase spontaneously breaks the twisted lattice translation symmetry.
There is a continuous phase transition between two gapped phases~\cite{Chen:2025ivo}. 
Similar to the previous example, we implement gauging to see the dual description of this DQCP. 
\paragraph{(i)~yAFM}
We first gauge $\mathbb Z_2^x$ in the yAFM Hamiltonian,
which is achieved by the standard KW duality. 
The gauged model reads
\begin{eqnarray}
    H^\prime_{\text{yAFM}}=-\sum_j\sigma^z_{j-1}\sigma^x_{j}\sigma_{j+1}^z.
\end{eqnarray}
In the dual model, the internal symmetry becomes 
\begin{equation}
    U_Q=\prod_j \sigma^x_{j},
    \qquad
    U_D=\prod_j (\sigma^{x}_{j})^j.
\end{equation}
Since the global Hadamard transformation does not commute with the \(\mathbb Z_2^x\) gauging map, it is promoted after gauging to a non-invertible defect obtained as~\cite{Cao:2024qjj}:
\begin{equation}
    \mathcal D_H
    =
    \mathcal D_{\mathrm{KW}}^\dagger
    \left(\prod_j H_j\right)
    \mathcal D_{\mathrm{KW}},\label{DH}
\end{equation}
where $D_{\mathrm{KW}}$ denotes KW duality operator regarding gauging internal $\mathbb Z_2^x$ symmetry. 
Hence, the twisted translation becomes the non-invertible lattice translation
\begin{equation}
    \widehat T'
    =
    T\,\mathcal D_H
    =
    T\,
    \mathcal D_{\mathrm{KW}}^\dagger
    \left(\prod_j H_j\right)
    \mathcal D_{\mathrm{KW}} .
\end{equation}
with fusion rule
\begin{equation}
    \widehat T'\times \widehat T'
    =
    \left(1+U_Q+U_D+U_QU_D\right)T^2 .\label{36}
\end{equation}
Besides the fusion rule, we evaluate the $F$-symbol~\eqref{Fsym} and show that $\widehat{T}'$ together with the internal symmetry forms a Rep($H_8$)-type symmetry. The details are relegated to Appendix.
\par
We further gauge the dipole symmetry, with the following spatially modulated Gauss law 
\begin{equation*}
    \tau^z_{j-\frac{1}{2}}(\sigma^x_j)^j\tau^z_{j+\frac{1}{2}}=1.
\end{equation*}
Accordingly, the gauged Hamiltonian is given by
\begin{eqnarray}\label{eq:gaugedyAFM}
       H^{\prime\prime}_{\text{yAFM}}=-\sum_{j}\sigma^z_{j-1}(\tau^x_{j-\frac{1}{2}})^{j-1}\sigma^x_j(\tau^x_{j+\frac{1}{2}})^{j-1}\sigma^z_{j+1}\label{afm},
\end{eqnarray}

Similar to the Rep$(D_8)$-type example, by dressing the translation operator with suitable local operators, one can construct a gauge-invariant non-invertible lattice translation, and then analyze its properties in the IR by imposing the Gauss law constraints. 
Moreover, because the gauging procedure is exactly the same as the previous case, we land in the same effective Hilbert space with effective degrees of freedom $\widehat{\sigma},\widehat{\tau}$ and the internal symmetry becomes $\widehat{\mathbb Z}_2^{\sigma}\times \widehat{\mathbb Z}_2^{\tau}$ generated by \eqref{eq:effectivesym}.
Therefore, we can construct the non-invertible lattice translation directly in the resulting effective Hilbert space by $\widehat{\mathcal{T}}'=\widehat{D}_{\tau}\widehat{D}_\sigma$, whose action on $\widehat{\sigma}$ spins is
\begin{align*}
    \widehat{\mathcal{T}}':\quad &\widehat{\sigma}^x_{k}\to \widehat{\sigma}^z_{k}\widehat{\sigma}^z_{k+1},\quad  \widehat{\sigma}^z_{k}\widehat{\sigma}^z_{k+1}\to \widehat{\sigma}^x_{k+1}.
\end{align*}
and similarly for $\widehat{\tau}$ spins. 
The fusion rule of this operator is identical to the Rep$(D_8)$-type one in~\eqref{fusionT}. 
However, this realizes a Rep($H_8$)-type crystalline symmetry by evaluation of the $F$-symbol following similar derivation in the Appendix. 

The whole gauging procedure is a crystalline analogue of the internal symmetry construction: for a purely internal $D_8$ anomaly, gauging either one $\mathbb{Z}_2$ subgroup or $\mathbb{Z}_2\times\mathbb{Z}_2$ symmetry yields Rep($H_8$) categorical symmetry~\cite{Diatlyk:2023fwf}. 
Here, the exotic LSM anomaly naturally yields a Rep($H_8$)-type non-invertible lattice translation after gauging.  
\par
In the effective space, the gauged Hamiltonian~\eqref{eq:gaugedyAFM} becomes
\begin{eqnarray}
    \widehat{H}_{\text{yAFM}}=-\sum_k\left(\widehat{\sigma}^z_{k}\widehat{\sigma}^z_{k+1}\widehat{\tau}^z_{k}\widehat{\tau}^z_{k+1}+\widehat{\sigma}^x_k\widehat{\tau}^x_k\right),
\end{eqnarray}
with two ground states
\begin{eqnarray*}
  \ket{\Omega_1}=  \bigotimes_k \left(\ket{0}_{\sigma,k}\ket{0}_{\tau,k}+\ket{1}_{\sigma,k}\ket{1}_{\tau,k}\right)\\
   \ket{\Omega_2}=  \bigotimes_k \left(\ket{0}_{\sigma,k}\ket{1}_{\tau,k}+\ket{1}_{\sigma,k}\ket{0}_{\tau,k}\right),
\end{eqnarray*}
spontaneously breaking $\mathbb{Z}_2^\sigma$ and $\mathbb{Z}_2^\tau$, yet preserving the diagonal subgroup of  $\mathbb{Z}_2^\sigma\times\mathbb{Z}_2^\tau$ and non-invertible translation.
Here $\ket{a}_{\sigma (\tau),k}(a=0,1)$ denotes diagonal basis of $\widehat{\sigma}^z(\widehat{\tau}^z)$ at site $k$.

\paragraph{(ii)~VBS}We turn to dual description of the VBS state which spontaneously breaks the twisted translation symmetry. The procedure of obtaining the gauged Hamiltonian is identical to the one in the previous argument around~\eqref{VBS2}-\eqref{27}. After coarse graining and projection, the gauged Hamiltonian is given by~\eqref{vbs5} with five ground states defined in~\eqref{27}. One can show that these ground states spontaneously break two internal symmetries as well as the non-invertible translation, and indicating that the dual VBS phase fully breaks Rep($H_8$)-type crystalline symmetry.\par
In summary, the yAFM–VBS transition admits an additional Landau description governed by Rep($H_8$)-type crystalline categorical symmetry, which is described by
\begin{equation}\label{eq:crystallandau2}
    \widehat{H}=(1-\lambda)\widehat{H}_{\text{yAFM}}+\lambda \widehat{H}_{\text{VBS}}.
\end{equation} 
Unlike the FM–VBS transition, which only admits a Rep($D_8$)-type description, the yAFM–VBS transition illustrates that a DQCP may admit multiple categorical descriptions distinguished by their categorical data.\par

\parR{Discussions and Outlook.}   
In this work, we have shown that LSM anomalies provide a microscopic mechanism for generating crystalline categorical symmetries through gauging, and that these symmetries can reorganize DQCPs into a Landau-type framework. In the examples studied here, gauging the anomalous onsite symmetry converts the original competition between magnetic and VBS orders into a transition between different symmetry-breaking patterns of a non-invertible crystalline symmetry. Thus, the apparent failure of the conventional Landau paradigm is not simply replaced by the absence of symmetry-breaking logic; instead, it is resolved by enlarging the notion of symmetry to include crystalline categorical symmetries. \par

A central lesson is that the categorical Landau description of a DQCP is captured by the full category data. The FM–VBS transition and the yAFM–VBS transition lead, after gauging, to non-invertible lattice translations with $\mathrm{Rep}$($D_8$)-type and $\mathrm{Rep}$($H_8$)-type categorical structures, respectively. These two categories share the same fusion rules but differ in their $F$-symbols. This distinction shows that the fusion ring alone is insufficient to characterize the categorical symmetry governing the transition. Instead, the associativity data of the category can have direct physical meaning in distinguishing categorical Landau descriptions of DQCPs.

Our construction also emphasizes the importance of working directly with microscopic lattice models. Instead of postulating a categorical symmetry in the IR limit, we derive the relevant microscopic non-invertible crystalline symmetries by gauging anomalous microscopic symmetries and track their action on concrete Hamiltonians and symmetry-breaking phases. This provides a lattice-level realization of categorical Landau transitions and clarifies how LSM constraints can serve as a bridge between conventional crystalline symmetries and generalized non-invertible symmetries. 

Our results suggest that categorical symmetry may provide a unifying language for deconfined critical phenomena. In this perspective, DQCPs are not merely exceptional violations of Landau theory, but can be viewed as Landau transitions for generalized, anomaly-generated symmetries. The natural order parameters are then not only conventional local operators, but also the symmetry-breaking patterns and associativity data of the underlying crystalline fusion category. Understanding this categorical structure would offer a more microscopic and symmetry-based route to organizing deconfined critical points and their possible universality classes.
\par

Several directions remain open:
\begin{enumerate}
    \item The yAFM-VBS transition has both ordinary and twist translations across the whole phase diagram, and is thus described by both Rep$(D_8)$- and Rep$(H_8)$-type Landau transition after the duality. It is interesting to find nontrivial DQCPs only described by dual Rep$(H_8)$-type crystalline symmetry.  Consider the following twisted VBS phase
\begin{equation}
    \begin{split}
      &H_{tVBS}=\sum_{k}(I-\sigma^x_{2k-1}\sigma^x_{2k})(I-\sigma^x_{2k}\sigma^x_{2k+1})\\
      &+\sum_{k}(I-\sigma^z_{2k}\sigma^z_{2k+1})(I-\sigma^z_{2k+1}\sigma^z_{2k+2}), \nonumber
    \end{split}
\end{equation}
which explicitly breaks the ordinary translation and the global Hadamard gate, yet preserves the combination of them -- the twist lattice translation symmetry.  
Therefore, the yAFM-tVBS transition (see Fig.\ref{fig:DQCPs}) only has twisted translation symmetry across the whole phase diagram, and is therefore a candidate example with only Rep($H_8$)-type crystalline Landau paradigm in the dual frame.
\item  A systematic classification of LSM anomalies that generate distinct crystalline categorical symmetries after gauging would help determine how broadly categorical Landau descriptions apply to DQCPs~\cite{PhysRevX.11.031043}.
\item It would also be valuable to extend the present construction beyond one spatial dimension, where crystalline symmetries, higher-form symmetries, and lattice anomalies can intertwine in richer ways~\cite{PhysRevB.109.245108,10.21468/SciPostPhys.20.4.117,Oishi:2026cvg}.
\end{enumerate}

\parR{Acknowledgments.}
We thank Masazumi Honda, Hosho Katsura, Linhao Li, Tsubasa Oishi, Takuma Saito, Qingrui Wang, Xueda Wen, Han Yan, and Ruizhen Huang for fruitful discussions. 
We acknowledge
support from 
JST CREST (Grant
No. JPMJCR24I3) (HE), Villum Fonden Grant no. VIL60714 (WC), the Deutsche Forschungsgemeinschaft (DFG, German Research Foundation) under Germany’s Excellence Strategy—Cluster of Excellence Matter and Light for Quantum Computing (ML4Q) EXC 2004/1 -- 390534769 as well as within the CRC network TR 183 (Project Grant No. 277101999) as part of subproject B01 (BH).\par
%
\parR{Note Added.}
While we were preparing this manuscript, we became aware of a recent work~\cite{Chen:2026fai} discussing spontaneous symmetry breaking patterns associated with $\mathrm{Rep}(H_8)$ symmetry and the corresponding phase transitions. We also became aware of an independent work~\cite{Zhang2026b} that studies DQCPs from the perspective of non-invertible symmetries using group cohomology argument. 
We thank the authors of Ref.~\cite{Zhang2026b} for coordinating the timing of our submissions.

\bibliography{main}
\begin{widetext}
\appendix
\section{A. Evaluation of the $F$-symbol}
In this section, following an approach presented in~\cite{PhysRevB.104.115156,Seifnashri:2023dpa,Seiberg:2024gek}, we evaluate the
$F$-symbol in Eq.~\eqref{Fsym} using the topological defect construction. To this end, we introduce defects associated with the invertible symmetries $\wZ_2^\sigma \times \wZ_2^\tau$
and the non-invertible defect $\mathcal{D} =S \wid{D}_\sigma \wid{D}_\tau$.
 We introduce a representative Hamiltonian preserving these symmetries:
\begin{eqnarray}
    H=-\sum_k(\wid{\sigma}^z_k\wid{\sigma}^z_{k+1}+\wid{\sigma}^x_k)+(\wid{\sigma}\leftrightarrow\wid{\tau}).\label{hamiltonianA}
\end{eqnarray}
We first consider inserting $\wZ_2^\sigma$ defect at the link between site $k=K-1$ and $k=K$, which is abbreviated as $(K-1,K)$~[In what follows, we denote a link which is connected with two adjacent sites, $k$ and $k+1$ as $(k,k+1)$]. The Hamiltonian~\eqref{hamiltonianA} is modified as
\begin{eqnarray}
    H^{(K-1,K)_\sigma}=-\sum_{k\neq K-1}(\wid{\sigma}^z_k\wid{\sigma}^z_{k+1}+\wid{\sigma}^x_k)
    +\wid{\sigma}_{K-1}^z\wid{\sigma}^z_K-\wid{\sigma}^x_K
    -\sum_k(\wid{\tau}^z_k\wid{\tau}^z_{k+1}+\wid{\tau}^x_k).
\end{eqnarray}
Equivalently, a $\wZ_2^\sigma$ defect can be realized by a twisted boundary condition, under which the $\wid{\sigma}^z$ variables on the two sides of the defect differ by a minus sign. \par
The $\wZ_2^\sigma$ defect is topological and can be
moved along the chain by a local unitary transformation. Indeed, the Hamiltonian with the defect  at the link $(K,K+1)$ is described by
\begin{eqnarray}
     H^{(K,K+1)_\sigma}=-\sum_{k\neq K}(\wid{\sigma}^z_k\wid{\sigma}^z_{k+1}+\wid{\sigma}^x_k)
    +\wid{\sigma}_{K}^z\wid{\sigma}^z_{K+1}-\wid{\sigma}^x_{K+1}
    -\sum_k(\wid{\tau}^z_k\wid{\tau}^z_{k+1}+\wid{\tau}^x_k),
\end{eqnarray}
which is written as
\begin{eqnarray}
  H^{(K,K+1)_\sigma} =  U_\sigma^{(K-1,K)} H^{(K-1,K)_\sigma}  U_\sigma^{(K-1,K)\dagger} 
\end{eqnarray}
with 
\begin{eqnarray}
     U_\sigma^{(K-1,K)}=\wid{\sigma}^x_{K}.
\end{eqnarray}
Inserting $\wZ_2^\tau$ defect is analogously discussed. 
\par
We also introduce insertion of the non-invertible defect $S\wid{D}_\sigma\wid{D}_\tau(\vcentcolon=\mathcal{D})$. The Hamiltonian in the presence of the defect at $(K-1,K)$ is described by
\begin{eqnarray}
     H^{(K-1,K)_{\mathcal{D}}}=-\sum_{k\neq K-1,K}(\wid{\sigma}^z_k\wid{\sigma}^z_{k+1}+\wid{\sigma}^x_k)-\sum_{k\neq K-1,K}(\wid{\tau}^z_k\wid{\tau}^z_{k+1}+\wid{\tau}^x_k)\nonumber\\
     -\wid{\sigma}^z_{K-1}\wid{\tau}^x_K-\wid{\tau}^z_{K-1}\wid{\sigma}^x_K-\wid{\sigma}^x_{K-1}-\wid{\tau}^x_{K-1}-\wid{\sigma}^z_K\wid{\sigma}^z_{K+1}-\wid{\tau}^z_K\wid{\tau}^z_{K+1}
\end{eqnarray}
\begin{figure}[t]
    \centering
\includegraphics[scale=0.9]{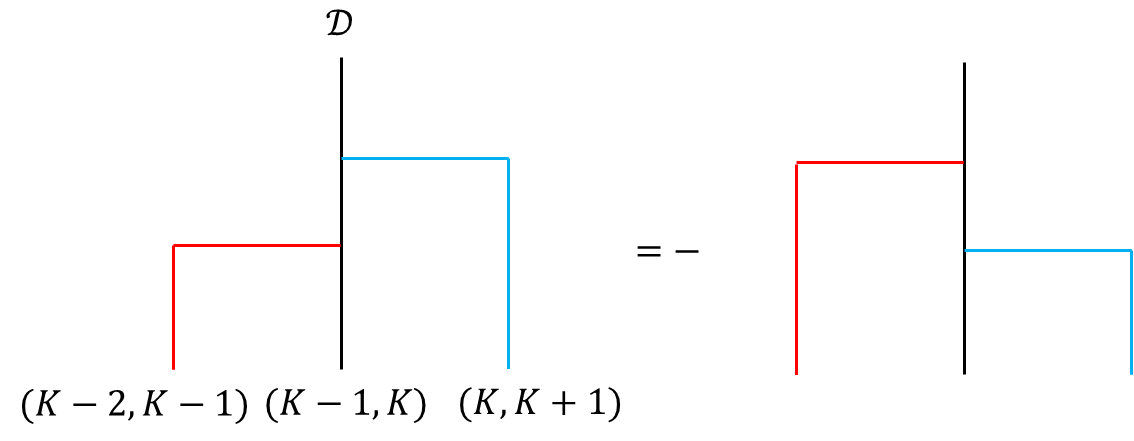}
    \caption{Graphical representation of the F-symbol. The red, black, and light-blue lines denote the $\wZ_2^\sigma$, $\mathcal{D}$, and $\wZ_2^\tau$ defects, respectively. The F-symbol compares two fusion paths, $(a\otimes \mathcal{D})\otimes b$ and $a\otimes(\mathcal{D}\otimes b)$, corresponding to opposite orders of fusing the invertible defects $a$ and $b$ into the non-invertible defect $\mathcal{D}$. In the lattice realization, the relative sign is determined by the commutation relation between the corresponding fusion operators. 
    }
    \label{fig:F}
\end{figure} 
Such a Hamiltonian can be constructed by implementing gauging at finite interval and change variables appropriately~\cite{Seiberg:2024gek}. Similar to the $\wZ_2^\sigma$ defect case, one can move the defect by a unitary operator. \par
We are now ready to evaluate the $F$-symbol. To do so, we introduce three defects associated with $\wZ_{2}^\sigma$, $\mathcal{D}$, and $\wZ_2^\tau$ at the link $(K-2,K-1)$, $(K-1,K)$, and $(K,K+1)$, respectively. 
Our goal is to extract the $F$-symbol by comparing two
different fusion paths of
the three defects. \par
The Hamiltonian in the presence of such defect configurations is described by
\begin{eqnarray}
    H^{(K-2,K-1)_\sigma,(K-1,K)_{\mathcal{D}},(K,K+1)_\tau}=
    -\sum_{k\neq K-2,K-1,K}(\wid{\sigma}^z_k\wid{\sigma}^z_{k+1}+\wid{\sigma}^x_k)-\sum_{k\neq K-1,K}(\wid{\tau}^z_k\wid{\tau}^z_{k+1}+\wid{\tau^x}_k)\nonumber\\
    -\wid{\sigma}^x_{K-2}+\wid{\sigma}^z_{K-2}\wid{\sigma}^z_{K-1}
     -\wid{\sigma}^z_{K-1}\wid{\tau}^x_K-\wid{\tau}^z_{K-1}\wid{\sigma}^x_K-\wid{\sigma}^x_{K-1}-\wid{\sigma}^z_{K}\wid{\sigma}^z_{K+1}-\wid{\tau}^x_{K-1}+\wid{\tau}^z_{K}\wid{\tau}^z_{K+1}.
\end{eqnarray}
The fusion of an invertible defect with \(\mathcal D\) has two steps: first move the invertible defect to the same link as \(\mathcal D\), and then apply the local fusion map. For example, we move the $\wZ_2^\sigma$ defect to the right to fuse $\mathcal{D}$: 
\begin{eqnarray}
    V_\sigma  H^{(K-2,K-1)_\sigma,(K-1,K)_{\mathcal{D}},(K,K+1)_\tau} V_\sigma^\dagger=  H^{(K-1,K)_{\mathcal{D}},(K,K+1)_\tau}
\end{eqnarray}
with 
\begin{eqnarray}
    H^{(K-1,K)_{\mathcal{D}},(K,K+1)_\tau}=-\sum_{k\neq K-1,K}(\wid{\sigma}^z_k\wid{\sigma}^z_{k+1}+\wid{\sigma}^x_k)-\sum_{k\neq K-1,K}(\wid{\tau}^z_k\wid{\tau}^z_{k+1}+\wid{\tau}^x_k)\nonumber\\
    -\wid{\sigma}^z_{K-1}\wid{\tau}^x_K-\wid{\tau}^z_{K-1}\wid{\sigma}^x_K-\wid{\sigma}^x_{K-1}-\wid{\tau}^x_{K-1}-\wid{\sigma}^z_K\wid{\sigma}^z_{K+1}+\wid{\tau}^z_K\wid{\tau}^z_{K+1}
\end{eqnarray}
and 
\begin{eqnarray}
      V_\sigma=\wid{\sigma}^x_{K-1}\wid{\tau}^z_K. 
\end{eqnarray}
We further move the $\wZ_2^\tau$ defect to the left to fuse $\mathcal{D}$:
\begin{eqnarray}
    V_\tau  H^{(K-1,K)_{\mathcal{D}},(K,K+1)_\tau}V_\tau=H^{(K-1,K)_{\mathcal{D}}}
\end{eqnarray}
with 
\begin{eqnarray}
    V_\tau=\wid{\tau}^x_K.
\end{eqnarray}
Comparing the two fusion paths shown in Fig.~\ref{fig:F} amounts
to comparing the actions of the fusion operators
$V_\sigma$ and $V_\tau$ in opposite orders. Since
\[
V_\sigma V_\tau = - V_\tau V_\sigma,
\]
the two fusion paths differ by a minus sign.
Therefore
\[
F^{D}_{aDb,D}=-1
\]
for \(a=(1,0)\) and \(b=(0,1)\).
By the same argument, one finds that the $F$-symbol is
trivial when $a$ and $b$ correspond to the same
$\mathbb Z_2$ factor.
These results establish that the two global symmetries
together with $\mathcal{D}$ realize Rep($D_8$) categorical symmetry.

\section{B. \texorpdfstring{$\mathrm{Rep}(H_8)$}{Rep(H8)}-type non-invertible lattice translation}

In this appendix, we determine the categorical structure
of the non-invertible twisted translation that emerges
after gauging the $U_X$ symmetry of the LSM-anomalous
system discussed in the main text~[around Eqs.~\eqref{DH}-\eqref{36}].
A subtle point is that the non-invertible Hadamard defect
$\mathcal{D}_H$ itself does not realize a Rep($H_8$)-type symmetry, where $\mathcal{D}_H$ is given in~\eqref{DH}.
Instead, $\mathcal{D}_H$ carries a Rep($D_8$)-type categorical
structure. Therefore, the Rep($H_8$) structure of the
gauged theory cannot be inferred from $\mathcal{D}_H$  alone.
Rather, one must include the lattice translation defect
and analyze the full non-invertible twisted translation
\begin{equation}
\hat T' = T \mathcal{D}_H .
\end{equation}
Our strategy is analogous to Appendix A.
We first review the categorical structure associated with
the Hadamard defect $\mathcal{D}_H$.
We then attach a translation defect and construct the
non-invertible twisted translation $\hat T'$.
Finally, by comparing different fusion processes, we
extract the corresponding lattice $F$-symbol and show that
the resulting bicharacter is precisely that of Rep($H_8$).

Before considering the twisted translation, let us first
review the categorical structure associated with the
Hadamard defect $\mathcal{D}_H$ itself.
Without loss of generality, start from the \(XX\) chain with the LSM anomaly and twisted translation,
\begin{equation}
    H=\sum_j \left(X_jX_{j+1}+Z_jZ_{j+1}\right).
\end{equation}
For notational simplicity, throughout this appendix we
use the conventional Pauli-operator notation
\(
X_j,Z_j
\)
instead of~\(
\sigma_j^x,\sigma_j^z
\).
After gauging \(U_X=\prod_j X_j\), the dual Hamiltonian becomes
\begin{equation}
    \widehat H=\sum_j \left(Z_{j-1}Z_{j+1}+X_j\right).
\end{equation}
with two global symmetries
\begin{align*}
    U_e=\prod_{j}X_{2j},\quad  U_o=\prod_{j}X_{2j-1}
\end{align*}

Throughout this appendix, we use the notation
$(U^e,U^o)$ 
instead of $(U_Q,U_D)$
to emphasize the correspondence with the
$\eta_e$ and $\eta_o$ defects.
The defect Hamiltonian with the defect of non-invertible Hadamard operator $\mathcal D_H$ inserted at link $(1,2)$ is~\cite{Cao:2024qjj}
\begin{equation}
    H_{\mathcal D_H}^{(1,2)}
    =
    Z_LX_l+X_1+Z_lZ_3+Z_1X_2
    +
    \sum_{j=3}^{L}
    \left(Z_{j-1}Z_{j+1}+X_j\right).
\end{equation}
It can be shown that, for even \(L\), the ordinary non-invertible defect \(\mathcal D_H\) alone has the \(\mathrm{Rep}(D_8)\)-type off-diagonal sign. Indeed, with an \(\eta^o\) defect on the left and an \(\eta^e\) defect on the right, one obtains fusion operators whose nontrivial parts include \(X_1Z_2\) and \(X_2\), and hence
\begin{equation}
    (X_1Z_2)X_2=-X_2(X_1Z_2),\label{55}
\end{equation}
The defect $\eta_e$ is represented by the fusion operator
$X_2$, while $\eta_o$ is represented by $X_1Z_2$.
The anticommutation relation~\eqref{55} therefore implies that
the fusion order of $\eta_e$ and $\eta_o$ differs by a
minus sign, giving
\[
\chi(\eta_e,\eta_o)=-1.
\]
Repeating the same analysis for the diagonal entries
yields
\[
\chi(\eta_e,\eta_e)=
\chi(\eta_o,\eta_o)=+1,
\]
which gives
\begin{equation}
\chi_{D_8}(a,b)
=
(-1)^{a_e b_o + a_o b_e}.
\end{equation}
Thus the non-invertible Hadamard defect $\mathcal{D}_H$ alone realizes a Rep($D_8$)-type categorical symmetry.
We now turn to the object of interest, namely the
non-invertible twisted translation
\(
\hat T' = T \mathcal{D}_H
\).
Unlike $\mathcal{D}_H$, its categorical structure receives a nontrivial contribution from the translation defect.
Therefore the resulting $F$-symbol need not coincide with
that of the Hadamard defect itself.\par
To compute the \(F\)-symbol for a non-invertible translation, we must also keep track of lattice translation defects. Following the construction in~\cite{Seiberg:2024gek}, a \(T^{-}\) defect removes one site, while a \(T^{+}\) defect inserts an additional site.
For example, a \(T^{-}\) defect is represented by
\begin{equation}
    H_{T^-}^{L}
    =
    -\left(Z_LZ_2+X_2\right)
    -
    \sum_{j=3}^{L}
    \left(Z_{j-1}Z_j+X_j\right),
\end{equation}
where the site \(1\) is effectively removed. A \(T^{+}\) defect at \((L,1)\) is represented by adding an intermediate site \((L,1)\):
\begin{align}
    H_{T^+}^{(L,1)}
    =
    -\left(Z_LZ_{(L,1)}+X_{(L,1)}\right)
    -
    \left(Z_{(L,1)}Z_1+X_1\right)
    -
    \sum_{j=1}^{L}
    \left(Z_{j-1}Z_j+X_j\right).
\end{align}
To fuse an internal defect with a translation defect, one first prepares the internal defect on a nearby link and then applies the appropriate movement operator to bring it into the translation defect. 
%

The defect of non-invertible twisted translation is obtained by fusing the \(\mathcal D_H\) defect with a translation defect. Start from the configuration with a $T^+$ defect and a $\mathcal{D}_H$ defect inserted at link $(L,1)$ and $(1,2)$ separately:
\begin{align}
    H_{T^+;\mathcal D_H}^{(L,1);(1,2)}
    =
    &Z_{L-1}Z_{(L,1)}
    +X_L
    +Z_LZ_1
    +X_{(L,1)}
   +
    Z_{(L,1)}X_l
    +X_1
    +Z_lZ_3
    +Z_1X_2
    +
    \sum_{j=3}^{L-1}
    \left(Z_{j-1}Z_{j+1}+X_j\right).
\end{align}
Conjugating by the movement operator~\cite{Cao:2024qjj}
\begin{equation}
    \lambda=S_{1,l}U^H_1\mathrm{CZ}_{1,2}
\end{equation}
gives the non-invertible twisted translation defect
\begin{align}
    H_{T'}^{(L,1)}
    =
    &Z_{L-1}Z_{(L,1)}
    +X_L
    +Z_LX_l
    +X_{(L,1)}
  +
    Z_{(L,1)}X_1
    +Z_lZ_2
    +Z_1Z_3
    +X_2
    +
    \sum_{j=3}^{L-1}
    \left(Z_{j-1}Z_{j+1}+X_j\right).
\end{align}

We now extract the bicharacter by comparing two representative fusions.

\paragraph{Off-diagonal entry: \texorpdfstring{$\eta^e$}{etae} on the left and \texorpdfstring{$\eta^o$}{etao} on the right.}

Place an \(\eta^e\) defect on the left of \(T'\) and an \(\eta^o\) defect on the right:
\begin{align}
   H_{\eta^e;T';\eta^o}^{(L-2,L-1);(L,1);(1,2)}
    =
    &-Z_{L-2}Z_L
    +X_{L-1}
    +Z_{L-1}Z_{(L,1)}
    +X_L
   +
    Z_LX_l
    +X_{(L,1)}
    +Z_{(L,1)}X_1
    +Z_lZ_2
    -Z_1Z_3
    \nonumber\\
    &+
    X_2
    +
    \sum_{j=3}^{L-1}
    \left(Z_{j-1}Z_{j+1}+X_j\right).
\end{align}
To move and fuse the left \(\eta^e\) defect into \(T'\), the relevant operator is
\begin{equation}
    X_LZ_l.
\end{equation}
To move and fuse the right \(\eta^o\) defect into \(T'\), the relevant operator is
\begin{equation}
    X_1.
\end{equation}
These two operators commute:
\begin{equation}
    (X_LZ_l)X_1
    =
    X_1(X_LZ_l).
\end{equation}
Since the corresponding fusion operators commute,
the two fusion paths are equivalent and no relative
phase is generated, that is, 
\begin{equation}
    \chi(\eta^e,\eta^o)=+1.
\end{equation}

\paragraph{Diagonal entry: \texorpdfstring{$\eta^e$}{etae} on both sides.}

Now place \(\eta^e\) defects on both sides of the non-invertible translation:
\begin{align}
    H_{\eta^e;T';\eta^e}^{(L-2,L-1);(L,1);(1,2)}
    =
    &-Z_{L-2}Z_L
    +X_{L-1}
    +Z_{L-1}Z_{(L,1)}
    +X_L
   +
    Z_LX_l
    +X_{(L,1)}
    +Z_{(L,1)}X_1
    +Z_lZ_2
    +Z_1Z_3
    \nonumber\\
    &+
    X_2
    -Z_2Z_4
    +X_3
    +
    \sum_{j=4}^{L-1}
    \left(Z_{j-1}Z_{j+1}+X_j\right).
\end{align}
The left \(\eta^e\) fusion again contributes
\begin{equation}
    X_LZ_l.
\end{equation}
whereas the right \(\eta^e\) fusion is represented by
\begin{equation}
    X_2X_l.
\end{equation}
These two operators anticommute:
\begin{equation}
    (X_LZ_l)(X_2X_l)
    =
    -(X_2X_l)(X_LZ_l)
\end{equation}
which implies that 
\begin{equation}
    \chi(\eta^e,\eta^e)=-1.
\end{equation}
By the same computation, or equivalently by
translation symmetry exchanging $\eta^e$ and $\eta^o$,
one finds
\begin{equation}
    \chi(\eta^o,\eta^o)=-1,
\end{equation}
Also, by the similar argument, we have
\begin{equation}
    \chi(\eta^e,\eta^o)
    =
    \chi(\eta^o,\eta^e)
    =
    +1.
\end{equation}
Hence, the full bicharacter is
\begin{equation}
    \chi(a,b)
    =
    (-1)^{a_e b_e+a_o b_o}.
\end{equation}
This is precisely the diagonal bicharacter of \(\mathrm{Rep}(H_8)\).

In conclusion, the non-invertible twisted translation obtained after gauging \(U_X\) has the fusion rule
\begin{equation}
    \widehat T'\times \widehat T'
    =
    \left(1+U^e+U^o+U^eU^o\right)T^2,
\end{equation}
and its lattice \(F\)-symbol is governed by the diagonal bicharacter
\begin{equation}
    \chi_{H_8}(a,b)
    =
    (-1)^{a_e b_e+a_o b_o}.
\end{equation}
Therefore, the gauged theory realizes a \(\mathrm{Rep}(H_8)\)-type non-invertible lattice translation, even though the unfused non-invertible Hadamard defect \(\mathcal D_H\) by itself carries the \(\mathrm{Rep}(D_8)\)-type off-diagonal bicharacter.

\end{widetext}
\end{document}